\documentclass{emulateapj}
\usepackage{amsmath}

\def\stacksymbols #1#2#3#4{\def\theguybelow{#2}
        \def\verticalposition{\lower#3pt}
        \def\spacingwithinsymbol{\baselineskip0pt\lineskip#4pt}
        \mathrel{\mathpalette\intermediary#1}}
\def\intermediary #1#2{\verticalposition\vbox{\spacingwithinsymbol
        \everycr={}\tabskip0pt
        \halign{$\mathsurround0pt#1\hfil##\hfil$\crcr#2\crcr
                \theguybelow\crcr}}}

\shorttitle{REMOVING COOL CORES AND CENTRAL METALLICITY PEAKS IN CLUSTERS}
\shortauthors{GUO \& MATHEWS}

\begin{document}
\bibliographystyle{apj} 

\title{Removing Cool Cores and Central Metallicity Peaks in Galaxy Clusters with Powerful AGN Outbursts}

\author{Fulai Guo\altaffilmark{1} and William G. Mathews \altaffilmark{1}}

\altaffiltext{1}{UCO/Lick Observatory, Department of Astronomy and Astrophysics, University of California, Santa Cruz, CA 95064, USA; fulai@ucolick.org}

\begin{abstract}
Recent X-ray observations of galaxy clusters suggest that cluster populations are bimodally distributed according to central gas entropy and are separated into two distinct classes: cool core (CC) and non-cool core (NCC) clusters. While it is widely accepted that AGN feedback plays a key role in offsetting radiative losses and maintaining many clusters in the CC state, the origin of NCC clusters is much less clear. At the same time, a handful of extremely powerful AGN outbursts have recently been detected in clusters, with a total energy $\sim 10^{61}-10^{62}$ erg. Using two dimensional hydrodynamic simulations, we show that if a large fraction of this energy is deposited near the centers of CC clusters, which is likely common due to dense cores, these AGN outbursts can completely remove CCs, transforming them to NCC clusters. Our model also has interesting implications for cluster abundance profiles, which usually show a central peak in CC systems. Our calculations indicate that during the CC to NCC transformation, AGN outbursts efficiently mix metals in cluster central regions, and may even remove central abundance peaks if they are not broad enough. For CC clusters with broad central abundance peaks, AGN outbursts decrease peak abundances, but can not effectively destroy the peaks. Our model may simultaneously explain the contradictory (possibly bimodal) results of abundance profiles in NCC clusters, some of which are nearly flat, while others have strong central peaks similar to those in CC clusters. A statistical analysis of the sizes of central abundance peaks and their redshift evolution may shed interesting insights on the origin of both types of NCC clusters and the evolution history of thermodynamics and AGN activity in clusters.

\end{abstract}


\section{Introduction}
\label{section:intro}

The hot gas in galaxy clusters emits prolifically in X-rays and has been extensively studied by X-ray telescopes {\it Chandra} and {\it XMM-Newton}, which reveal a striking bimodality in the properties of cluster cores. According to their core gas entropy, clusters are separated into two distinct classes: cool core (CC) and non-cool core (NCC) clusters. While the former have a low central gas entropy peaked at $S\equiv
k_{\rm B}T/ n_{\rm{e}}^{2/3} \sim 15 \, {\rm keV \, cm^{2}}$, the latter usually have high-entropy cores peaked at $S \sim 150 \, {\rm keV \, cm^{2}}$, and there is a distinct less-populated gap between $\sim 30$ and $\sim 50\, {\rm keV \, cm^{2}}$ \citep{cavagnolo09}. This bimodality also appears in cluster temperature profiles, which decrease significantly toward the center in CC clusters, but remain relatively flat in NCC clusters \citep{sanderson06}. Observations also suggest that AGN activity and star formation in cluster central dominant galaxies are much more pronounced in CC clusters \citep{burns90,cavagnolo08, rafferty08}.

The origin of this dichotomy has been studied by many authors. Galaxy clusters may naturally reach the CC state due to the interplay between radiative cooling and heating by active galactic nuclei (AGNs). It is widely thought that AGN outbursts play a key role in heating the intracluster medium (ICM), thus preventing cooling catastrophe (see \citealt{mcnamara07} for a recent review). It was also shown that AGN may operate as a self-regulating feedback mechanism, which is essential in suppressing global thermal instability and thus in maintaining the ICM in the CC state \citep{guo08b}. In contrast, the origin of NCC clusters is much less clear; three competing explanations have been proposed:
\begin{enumerate}
  \item Mergers: Mergers are shown to significantly disturb cluster CCs and mix the ICM \citep{ricker01,gomez02,ritchie02}. However, recent simulations by \citet{poole08} find that CC systems are remarkably robust and can be disrupted only in direct head-on or multiple collisions; even so, the resulting warm core state is only transient. Cosmological simulations by \citet{burns08} suggest that NCC clusters may form when they experience major mergers early in their evolution which destroy embryonic CCs. Note that clusters in these simulations suffer from the overcooling problem since they do not incorporate mechanisms (such as AGN feedback) to stop cooling catastrophe. Furthermore, the relatively low numerical resolution in these cosmological simulations ($15.6 \, h^{-1} {\rm kpc}$) may preclude firm conclusions about core structure and evolution.
    
 \item  Pre-heating: \citet{mccarthy08} suggested that early pre-heating prior to cluster collapse could explain the formation of  CC/NCC systems, which receive lower/higher levels of pre-heating. A possible concern in such scenarios is whether one can pre-heat the ICM to a high adiabat and yet retain sufficient low entropy gas in lower mass halos to obtain a realistic galaxy population. Furthermore, observations by \citet{rossetti10} find that most NCC clusters host regions with low-entropy, but relatively high metallicity gas, which are probably a signature of recent CC to NCC transformation, apparently inconsistent with both the scenarios proposed by \citet{mccarthy08} and \citet{burns08}.
 
  \item AGN outbursts: Strong AGN outbursts with $E_{\rm agn}$ $\sim 10^{61}$-$10^{62}$ erg have been detected in clusters, e.g., Hydra A \citep{nulsen05a}, Hercules A \citep{nulsen05} and MS0735.6+7421 (\citealt{mcnamara05}; \citealt{mcnamara09}). Recently in \citet[hereafter GO09]{guo09}, we show that such strong AGN outbursts could bring a CC cluster to the NCC state, which can be stably maintained by conductive heating from the cluster outskirts. AGN outbursts may also drag magnetic field lines radially, thus enhancing thermal conduction, which could significantly heat CCs in high-temeprature clusters.
 \end{enumerate}

The origin of NCC clusters may not necessarily explain the origin of the CC/NCC bimodality. To ensure the pronounced bimodality seen in observations, clusters need to stay in the NCC state for a duration at least comparable to the cooling time. In fact, many NCC clusters have a short cooling time ($\sim 1$ Gyr; \citealt{sanderson06}), and the NCC peak in the cluster central entropy distribution may not exist unless heating largely offsets radiative cooling for a relatively long time. While clusters may be maintained in the CC state by episodic AGN feedback, thermal conduction can stably keep clusters in the NCC state \citep{guo08b}. This has been confirmed by numerical simulations in GO09, which further predicts that the CC/NCC dichotomy is more pronounced in higher-temperature clusters, due to the fact that the heating and stablizing effects of conduction decline with temperature. 

In GO09, we adopted a simplified `effervescent model' for AGN heating \citep{begelman01}, and performed one-dimensional (1D) calculations. X-ray cavities in this model are in pressure equilibrium with and rise buoyantly in the ICM, resulting in a very gentle AGN heating. Furthermore, the 1D model assumes spherical symmetry. However, real AGN outbursts produce jets and X-ray cavities in two opposite directions, which are by no means spherically symmetric. The formation and evolution of X-ray cavities are probably much more dynamic than assumed in the `effervescent model'. Shock waves have been detected in many clusters with X-ray cavities, and are usually thought to be a natural result of AGN energy released in the ICM. In this paper, we study the evolution of X-ray cavities formed as AGN energy is injected into the ICM, using two-dimensional (2D) hydrodynamical simulations. Our primary goal is to investigate if strong AGN outbursts can transform a CC cluster to the NCC state, and if so, how this transformation happens.

The metallicity distribution of the ICM, extensively measured with X-ray observations, contains important clues about the physics of galaxy clusters. The mass and distribution of metals in the ICM constrain the integrated history of past star formation (metals are released via supernovae explosions and winds) and the ICM enrichment processes (see \citealt{bohringer10} for a recent review). The spatial abundance distribution is also significantly affected by transport processes in the ICM, e.g., turbulent mixing triggered by central AGN outbursts \citep{rebusco05, rebusco06, roediger07}. Of great interest is the bimodality in central abundance profiles between CC and NCC clusters. CC clusters usually have a peak in the iron abundance profile at the cluster center, while many NCC clusters show a relatively flat radial abundance profile \citep{degrandi01, degrandi04,leccardi10}. Nevertheless, recent observations by \citet{leccardi08} and \citet{sanderson09} suggest that some NCC clusters also have central iron abundance peaks. The differences in abundance profiles between CC and NCC systems have not been studied with detailed computational models. If NCC clusters are transformed from CC systems (i.e., not primordial), the same process transforming the CC to NCC systems may also explain the differences seen in their abundance profiles. In this paper we follow the evolution of iron abundance in our simulations and investigate if strong AGN outbursts can significantly change abundance profiles as they transform a CC cluster to the NCC state. We show that strong AGN outbursts efficiently mix metals in cluster central regions, but the final abundance profile in the NCC state strongly depends on the size of the initial abundance peak at the CC cluster center. Our model may naturally explain the the range of abundance profiles observed in NCC clusters. CC systems with broad/narrow central abundance peaks can be transformed by powerful AGN outbursts into NCC systems with/without central abundance peaks. A statistical analysis of the sizes of abundance peaks and their redshift evolution could test the validity and applicability of our model and may shed interesting insights on the origin of the CC/NCC bimodality.
 
The rest of the paper is organized as follows. In
Section~\ref{section2}, we present our model, including basic equations and numerical setup. Our results are presented in Section~\ref{section:results}. We summarize our main results in
Section~\ref{section:conclusion} with a discussion of the implications. The cosmological parameters used throughout this paper are: $\Omega_{m}=0.3$, $\Omega_{\Lambda}=0.7$, $h=0.7$. We have rescaled observational results if the original paper used a different cosmology.
 
\section{Equations and Numerical Methods}
\label{section2}

\subsection{Basic Equations}
\label{section:equation}

AGN outbursts inject cosmic rays (CRs) into the ICM, producing X-ray cavities in the cluster gas which have also been observed at radio frequencies in many clusters due to synchrotron emission of relativistic electrons. The outbursts induce weak shocks, and drive large amounts of gas mass outflow \citep{guo10}. In this paper, we study the combined evolution of thermal gas, CRs, and the gas metallicity during and after outbursts.
The governing equations may be written as:

\begin{eqnarray}
\frac{d \rho}{d t} + \rho \nabla \cdot {\bf v} = 0,\label{hydro1}
\end{eqnarray}
\begin{eqnarray}
\rho \frac{d {\bf v}}{d t} = -\nabla (P+P_{\rm c})-\rho \nabla \Phi ,\label{hydro2}
\end{eqnarray}
\begin{eqnarray}
\frac{\partial e}{\partial t} +\nabla \cdot(e{\bf v})=-P\nabla \cdot {\bf v}-n_{\rm i}n_{\rm
  e}\Lambda(T,Z)
   \rm{ ,}\label{hydro3}
   \end{eqnarray}
\begin{eqnarray}
\frac{\partial e_{\rm c}}{\partial t} +\nabla \cdot(e_{\rm c}{\bf v})=-P_{\rm c}\nabla \cdot {\bf v}+\nabla \cdot(\kappa\nabla e_{\rm c})+\dot{S_{\rm c}}
   \rm{ ,}\label{hydro4}   \end{eqnarray}
\begin{eqnarray}
\frac{d \rho_{\rm Fe}}{d t} + \rho_{\rm Fe} \nabla \cdot {\bf v} = 0,\label{hydro5}
\end{eqnarray}   \\ \nonumber
\noindent
where $d/dt \equiv \partial/\partial t+{\bf v} \cdot \nabla $ is the
Lagrangian time derivative, $P_{\rm c}$ is the CR pressure, $e_{\rm
  c}$ is the CR energy density, $\kappa$ is the CR diffusion
coefficient, $\dot{S_{\rm c}}$ is the CR source term due to the
central AGN activity, $\rho_{\rm Fe}$ is the iron density, and all other variables have their usual
meanings. Pressures and energy densities are related via
$P=(\gamma-1)e$ and $P_{\rm c}=(\gamma_{\rm c}-1)e_{\rm c}$, where we
assume $\gamma=5/3$ and $\gamma_{\rm c}=4/3$.  

Equation (\ref{hydro5}) describes the conservation of iron mass. Since we focus on the effect of AGN outbursts on the iron distribution and follow the cluster evolution for a timescale much shorter than enrichment times ($\gtrsim 5$ Gyr; \citealt{bohringer04}), we ignore the iron source term. The iron abundance $Z$ in units of the solar value is proportional to $Z\propto \rho_{\rm Fe}/\rho$. Thus the iron density $\rho_{\rm Fe}$ in equation (\ref{hydro5}) may be replaced by $Z \rho$. Since both the gas mass and iron mass are conserved, the metallicity $Z$ is also conserved: 
\begin{eqnarray}
\frac{dZ}{dt} \equiv  \frac{\partial Z}{\partial t} +  {\bf v}\cdot \nabla Z   = 0{\rm ,}
\end{eqnarray}
and therefore provides a tracer for the ICM gas, which is helpful in understanding how AGN outbursts affect and mix the ICM.

In the gas energy equation (\ref{hydro3}), we include radiative cooling with a volume cooling rate
$n_{\rm i}n_{\rm e}\Lambda(T,Z)$, where the cooling function $\Lambda(T,Z)$ is adopted from \citet{sd93}
and depends on both gas temperature $T$ and metallicity $Z$. The ion number density $n_{\rm i}$ is related to the proton number density $n_{\rm H}$ via $n_{\rm i}=1.1n_{\rm H}$, and thus the molecular weight is $\mu=0.61$. The gas temperature is related to the gas pressure and density via the ideal gas law:
\begin{eqnarray}
T=\frac{\mu m_{\mu}P}{k_{\rm B} \rho} {\rm ,}
   \end{eqnarray}
where $k_{\rm B}$ is Boltzmann's constant and $m_{\mu}$ is the atomic mass unit.

Equations (\ref{hydro1}) $-$ (\ref{hydro5}) are solved in $(r, z)$
cylindrical coordinates using a two-dimensional Eulerian code similar
to ZEUS 2D \citep{stone92}; in particular, we have incorporated into
the code a background gravitational potential, CR diffusion, CR
energy equation, and iron equation of mass conservation. The computational grid consists of $100$ equally
spaced zones in both coordinates out to $100$ kpc plus additional
$100$ logarithmically-spaced zones out to $1$ Mpc. 
For all the three fluids, we adopt reflective
boundary conditions at the origin and outflow boundary conditions at 
the outer boundary.

\subsection{Initial Cluster Profiles}
\label{section:ic}

 \begin{figure}
\plotone{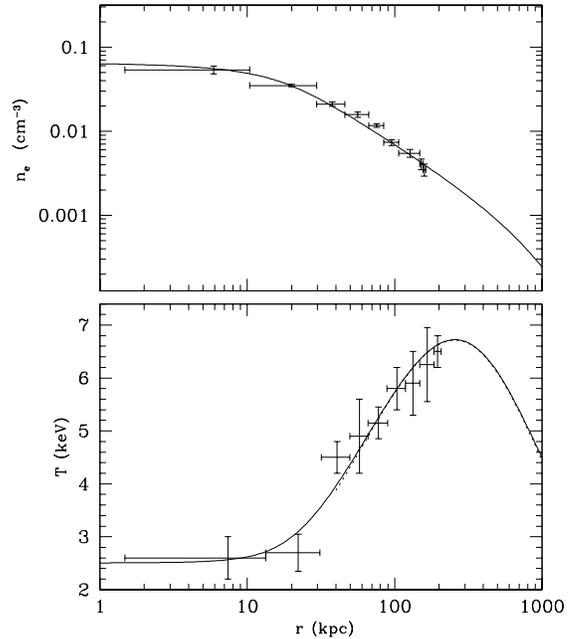}
\caption{Initial radial profiles of electron number density (top panel) and temperature (bottom panel). The dotted line in the bottom panel is the best analytical fit to Chandra data between 40 to 1000 kpc by \citet{vikhlinin06}; crosses in both panels indicate Chandra data by \citet{ettori02}. Our initial cluster profiles (solid lines) provides a very good fit to data for the whole cluster out to $1$ Mpc.}
 \label{plot1}
 \end{figure}

\begin{figure}
\plotone{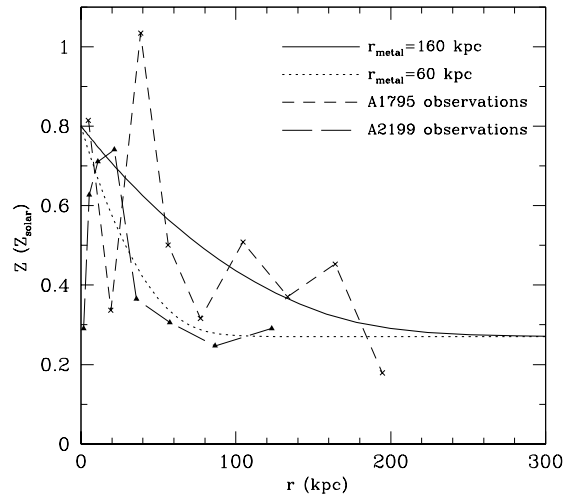}
\caption{Initial abundance profiles. The solid line (adopted in our main run D1) has a central abundance peak with size $r_{\rm metal}=160$ kpc and fits Chanda data of A1795 (crosses; \citealt{ettori02}) reasonably well, which have large scatter. The dotted line has a much smaller central abundance peak ($r_{\rm metal}=60$ kpc) and is adopted in run D2. Chandra data for the cluster A2199 (triangles; \citealt{johnstone02}), which has a much smaller central abundance peak, is shown for comparison.}
 \label{plot2}
 \end{figure}
 
Our model and methods are generally applicable to all relaxed clusters,
but for concreteness, we adopt simulation parameters appropriate for the typical CC cluster
Abell 1795, which has been well observed by both {\it Chandra} and {\it XMM-Newton} \citep{tamura01, ettori02, vikhlinin06}.

For initial profiles of A1795, we first build an analytic fit to the deprojected $3$-dimensional gas temperature profile derived from {\it Chandra} observations, which acquired data out to $\sim 1$ Mpc covering our entire computational grid:
\begin{eqnarray}
T(r)=(T_{\rm in}^{\alpha}+T_{\rm V}^{\alpha})^{1/\alpha}{,}
\end{eqnarray}
where the constant $\alpha$ is chosen to be $\alpha=5$, $T_{\rm in}=2.5$ keV is the observed central temperature of A1795, and $T_{\rm V}$ is the best-fit temperature profile of \citet{vikhlinin06} which provides an excellent fit to {\it Chandra} data of A1795 from $40$ kpc to $1$ Mpc:
\begin{eqnarray}
T_{\rm V}(r)=9.68t_{\rm cool}(r)t_{\rm out}(r)\text{ keV,}
\end{eqnarray}
where 
\begin{eqnarray}
t_{\rm cool}(r)=\frac{0.1+(r/r_{1})^{1.03}}{1+(r/r_{1})^{1.03}}\text{ ,}
\end{eqnarray}
describes that the temperature declines inward in the central region, and
\begin{eqnarray}
t_{\rm out}(r)=\frac{1}{[1+(r/r_{2})^{a}]^{b/a}}\text{ ,}
\end{eqnarray}
represents the radially declining outer region. Here the parameters are $r_{1}=77$ kpc, $r_{2}=550$ kpc, $a=1.63$ and $b=0.9$. The resulting temperature profile $T(r)$ is shown as the solid line in the bottom panel of Figure \ref{plot1}, where $T_{\rm V}(r)$ (the dotted line) and {\it Chandra} data (crosses) from \citet{ettori02} are also shown. Clearly our analytic temperature profile provides a very good fit to observations out to $\sim 1$ Mpc.

The initial cluster density profile is solved by assuming hydrostatic equilibrium.
At the beginning of our simulation $t=0$, the CR energy density is assumed to be $e_{\rm cr}=0$ throughout the cluster. The gravitational potential $\Phi$ is determined by the dark matter Navarro-Frenk-White (NFW) profile \citep{navarro97}:
\begin{eqnarray}
\rho_{\text{DM}}(r)=\frac{M_{0}/2\pi}{r(r+r_{\text{s}})^{2}}\text{,}  
\end{eqnarray}
where $r_{\text{s}}$ is  the standard scale radius  of the NFW profile and $M_{0}$ is a characteristic mass. This density distribution results in a gravitational potential:
\begin{eqnarray}
\Phi= -\frac{2GM_{0}}{r_{\text{s}}} \frac{\text{ln}(1+r/r_{\text{s}})}{r/r_{\text{s}}}    \text{.} 
\end{eqnarray}
For A1795, we choose $M_{0}=4.8\times 10^{14}M_{\odot}$, and $r_{\text{s}}=460$ kpc. At redshift $0.0632$ \citep{ettori02}, these parameters correspond to a virial mass $M_{\rm vir}=7.87\times 10^{14}M_{\odot}$, and a concentration $c=4.07$ assuming that the virial radius is $r_{\rm vir}\sim r_{200}$, within which the mean mass density is $200$ times the critical density of the Universe. Taking the electron number density $n_{\rm e}$ at the origin to be $n_{0}=0.065$ cm$^{-3}$, we derive from hydrostatic equilibrium the initial radial profile of $n_{\rm e}$, which fits observations very well, as clearly shown in the {\it top} panel of Figure \ref{plot1}. 

X-ray observations have been extensively used to measure abundance profiles in galaxy groups and clusters. 
In CC clusters, the observed abundance distributions typically peak in the central, low-entropy core regions, and decline radially outward \citep{degrandi01, degrandi04, baldi07,leccardi08}. These observations also suggest that abundance profiles flatten off with $Z\sim 0.2-0.3$ at large radii. In this paper the metallicity $Z$ is calculated relative to the solar iron abundance ($Z_{\odot}=$Fe/H$=4.68\times 10^{-5}$ by number) published by \citet{anders89}. It has been superseded by the new value $Z_{\odot}=3.16\times 10^{-5}$ of \citet{grevesse98} and \citet{asplund09}, but it allows straightforward comparison with most of the literature. Furthermore, the metallicity $Z$ used in the \citet{sd93} cooling function $\Lambda(T,Z)$ is also in units of the \citet{anders89} iron abundance. A simple scaling by $1.48$ converts our abundance values to the metalicities relative to the new \citet{grevesse98} solar abundance, without changing any other result in our paper.

The initial abundance profile is chosen to be
\begin{eqnarray}
Z=(Z_{0}^{\beta}+Z_{\rm r}^{\beta})^{1/\beta}
\text{,}   \label{metaleq} 
\end{eqnarray}
where $\beta=5$. The constant $Z_{0}=0.27$ sets a minimum for the abundance profile, which represents the flattening off at large radii, and is taken from \citet{degrandi04}, where we have converted metalicities in unit of the \citet{grevesse98} iron abundance to those relative to the \citet{anders89} abundance. 
The outward abundance decline is described by
\begin{eqnarray}
Z_{\rm r}=0.8e^{-r/r_{\rm metal}}
\text{,}  \label{metalr} 
\end{eqnarray}
where $r_{\rm metal}$ characterizes the spatial size of the central metallicity peak. The initial abundance profiles for our runs and {\it Chandra} data are shown in Figure \ref{plot2}.

\subsection{Cosmic-ray Physics and Assumptions}
\label{section:cr}

X-ray cavities are usually thought to be inflated by bipolar jets emanating from AGNs located at cluster centers. 
Radio synchrotron emission has been detected from many X-ray cavities (especially when the cavities are young), confirming the existence of a significant amount of CRs, which may be transported along the jets and/or created in strong shocks as the jets encounter the ICM. One can envision that the jets deposit CRs into small regions at their terminal points, which expand and form underdense bubbles producing the observed X-ray cavities. We have thus developed a model to numerically study CR injection and evolution in the ICM, which has been successfully used in \citet{mathews08a}, \citet{mathews08}, \citet{mathews09}, and \citet{guo10}, where the reader is referred to for further details. Here we simply summarize several modifications and reiterate
some important points. 

The injection of CRs into the ICM is described 
in equation (\ref{hydro4}) by the
source term $\dot{S_{\rm c}}$. We assume that the CRs are
deposited into a Gaussian-shaped sphere of characteristic radius
$r_{\rm s}=5$ kpc located at ${\bf r}_{\rm cav}=(r,z)=(0, z_{\rm
  cav})$:
\begin{eqnarray}
\dot{S_{\rm c}}= 
\begin{cases}
\frac{E_{\rm agn}}{t_{\rm agn}}\frac{e^{-[({\bf r}-{\bf r}_{\rm cav})/r_{\rm s}]^{2}}}{\pi^{3/2}r_{\rm s}^{3}}      & \quad \text{when $t \leq t_{\rm agn}$,}\\ 
0& \quad \text{when $t>t_{\rm agn}$,}
\end{cases}
\end{eqnarray}
\noindent
where $t_{\rm agn}=10$ Myr is the duration of the CR injection (AGN active
phase), and $E_{\rm agn}$ is the total injected CR energy in one
bubble ($2E_{\rm agn}$ for the whole cluster). The integral of
$\dot{S_{\rm c}}$ over space gives the the CR injection luminosity
$E_{\rm agn}/t_{\rm agn}$ during the active phase. See Table \ref{table1} for specific model parameters in each run.

In addition to their advection with the thermal gas, CRs
diffuse through the gas as 
described in equation (\ref{hydro4}). The CR diffusion coefficient
$\kappa$ is poorly known but may vary inversely with the gas
density since the magnetic field is probably larger in denser gas
\citep{dolag01}. We adopt the following functional dependence of the diffusion
coefficient on the gas density:
\begin{eqnarray}
\kappa=
\begin{cases}
10^{30}(n_{{\rm e}0}/n_{\rm e}) \text{~cm}^{2} \text{~s}^{-1}    & \quad \text{when } n_{\rm e}> n_{{\rm e}0} \text{,}\\
10^{30} \text{~cm}^{2} \text{~s}^{-1}    & \quad \text{when } n_{\rm e} \leq n_{{\rm e}0} \text{,}
\end{cases}
\end{eqnarray}
\noindent
where $n_{{\rm e}0}=10^{-5}$ cm$^{-3}$. This level of diffusion does not strongly affect the early evolution of cavities,
and our results are fairly insensitive to it. 
During their diffusion, CRs interact with
magnetic irregularities and Alfv\'{e}n waves, exerting CR pressure
gradients on the thermal gas (equation \ref{hydro2}). 
The early evolution of cavities and associated shock heating are mainly determined by 
the expansion of the surrounding gas driven by the CR pressure.
We neglect other more complicated (probably secondary) interactions of CRs with thermal gas, e.g.,
Coulomb interactions, hadronic collisions, and
hydromagnetic-wave-mediated CR heating, that depend on the CR
energy spectrum and provide additional heating effects for the ICM
(e.g., \citealt{guo08a}). Our model reproduces very well many characteristic morphological 
features of the huge X-ray cavities observed in the cluster MS 0735.6+7421 \citep{guo10} 
which experienced an AGN energy release comparable to that considered here.

\begin{table}
 \centering
 \begin{minipage}{70mm}
  \renewcommand{\thefootnote}{\thempfootnote} 
  \caption{List of Simulations}
    \vspace{0.1in}
  \begin{tabular}{@{}lccccc}
  \hline & {$E_{\rm agn}$\footnote{The AGN energy released in the form
      of CRs in one hemisphere during the active phase $t\leq t_{\rm agn}=10$ Myr.
      } } & {$z_{\rm cav}$\footnote{${\bf r}_{\rm
               cav}=(0, z_{\rm cav}$) is the position where the cosmic
             rays are injected during the AGN active phase.}} &
         {$r_{\rm metal}$\footnote{$r_{\rm metal}$ characterizes the spatial size of the central abundance peak (see equation \ref{metalr}).}} & {$r_{0}$\footnote{$r_{0}$ is an inner cutoff radius characterizing a central dip in the initial cluster abundance profile, which is only introduced in run D1-D (see Section~\ref{section:rcap}). }}\\ Run&
         $(10^{61}$ erg)& (kpc)&(kpc)&(kpc)\\ \hline D1 &3.15&10& 160 &- \\ D1-A & 3.15& {10-50\footnote{In this run, $z_{\rm cav}$ moves from 10 to 50 kpc with a constant speed during the CR injection phase.}} & 160&- \\ D1-B & 6.3&
         10 & 160 &- \\ D1-C &3.15 & 30 & 160 &- \\D1-D &3.15 & 10 & 160 &10 \\D2 &3.15 & 10 & 60 &- \\D2-A &3.15 & 10 & 40 &- \\
         
          \hline
\label{table1}
\end{tabular}
\end{minipage}
\end{table}

\section{Results}
\label{section:results}

\subsection{Removal of Cool Cores}
\label{section:rcc}

\begin{figure}
\plotone{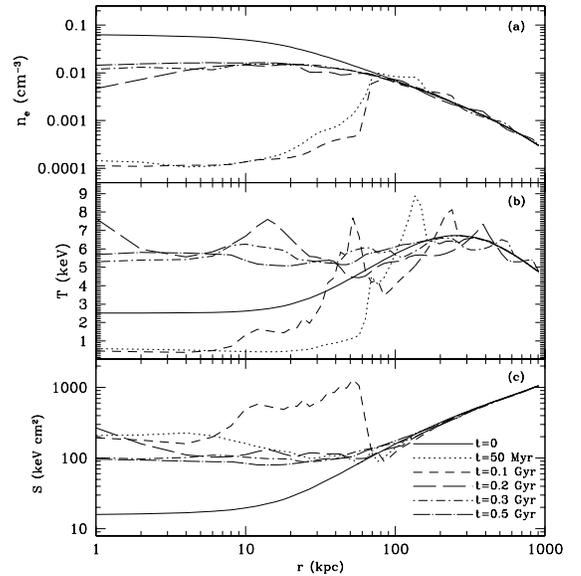}
\caption{Time evolution of radial profiles of emission-weighted spherically averaged (a) electron number density, (b) temperature, and (c) entropy in our main run D1. After $t\sim 0.2-0.3$ Gyr, the cluster relaxes to the NCC state with a relatively flat temperature profile and a high central entropy ($\sim 100$ keV cm$^{2}$) core.}
 \label{plot3}
 \end{figure}
 
\begin{figure*}
\plotone{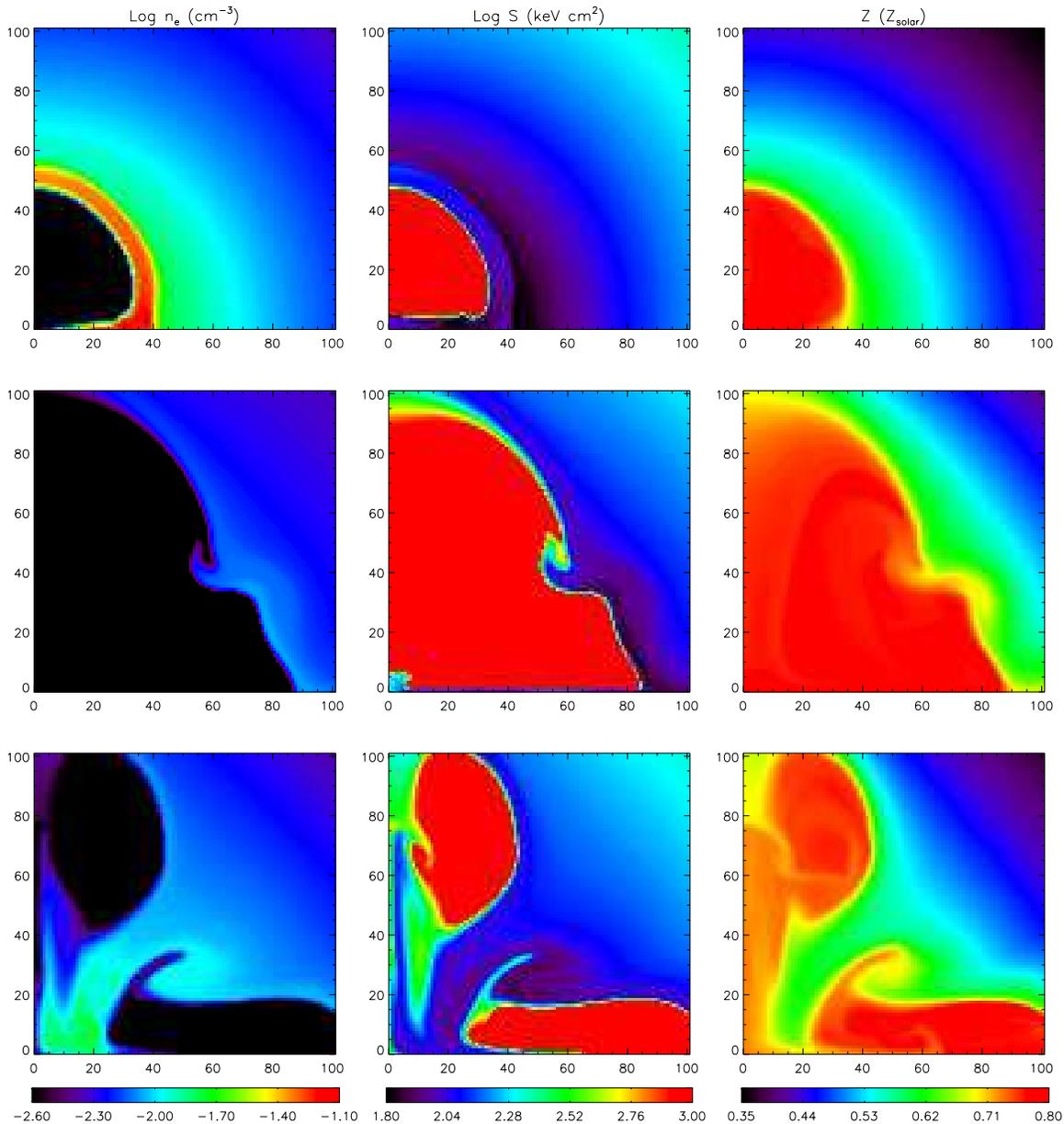}
\caption{Central slices ($100\times100$ kpc) of  log $(n_{\rm e}/{\rm cm}^{-3})$ (left column), log $(S/{\rm keV~ cm}^{2})$ (middle column), and  log $(Z/Z_{\rm solar})$ (right column) at three time epochs t=$10$ (top), $100$ (middle), $200$ Myr (bottom) in run D1. The horizontal and vertical axes represent $r$ and $z$ in kpc respectively. Cavity regions in the left and middle panels are saturated to show the variations of gas density and entropy in the ICM. The images clearly show the formation and final break-up of the low-density cavity produced by the AGN outburst.}
 \label{plot4}
 \end{figure*}
  
\begin{figure}
\plotone{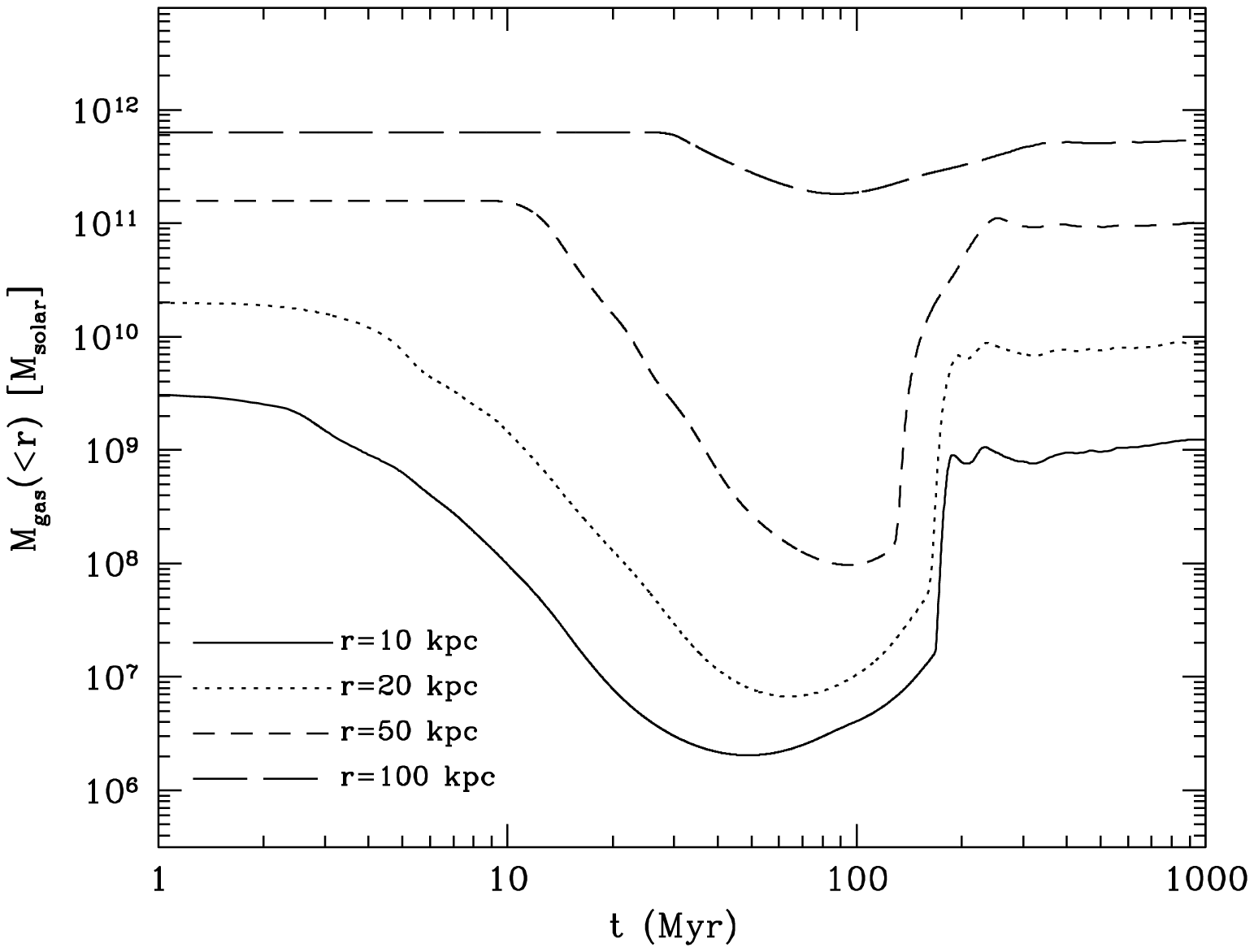}
\caption{Time evolution of gas mass $M_{\rm gas}(<r)$ enclosed within various radii $r=10,20,50, 100$ kpc in run D1. The AGN outburst deposits CRs into the ICM, which cause a large gas outflow, creating a low-density cavity at the cluster central region. At $t\sim 0.1-0.2$ Gyr, the cavity starts to break up as thermal gas falls back to the center.}
 \label{plot5}
 \end{figure}
 
The primary goal of this paper is to determine if powerful AGN outbursts can transform CC clusters to the NCC state, and if they can, how central abundance peaks typically found in CC clusters are affected during this transformation. We begin with the removal of cool cores in this subsection. Our model is intended to be generic, but for definiteness, we choose the CC cluster A1795 as our fiducial cluster. Model parameters in our runs are listed in Table \ref{table1}. We first present the results of our main run D1, and then compare this with runs computed with other combinations of the parameters in Table 1.

Figure \ref{plot3} shows time evolution of radial profiles of emission-weighted spherically averaged gas properties, including (a) electron number density, (b) temperature, and (c) entropy in run D1. The solid lines represent the observed profiles of the CC cluster A1795, which serve as the initial state in our simulations. Within the CC ($\lesssim 100$ kpc), both the temperature and entropy decrease inward and reach low values $T\sim 2.5$ keV, and $S\sim 10$-$20$ kev cm$^{2}$ at the cluster center. As AGN energy in the form of CRs is injected into the ICM, shocks appear and heat the gas, as clearly seen in the top-left panel of Figure \ref{plot4} (the shock front is located near the outer edge of the red `annulus'), which shows central slices ($100\times100$ kpc) of electron number density, entropy, and metallicy distributions at three times t=$10$ (top panels), $100$ (middle panels), $200$ Myr (bottom panels). The dotted line in Figure \ref{plot3}(c) suggests that all the gas in the CC has already been heated to a high entropy $\sim 100$-$200$ kev cm$^{2}$ at $t=50$ Myr, but Figure \ref{plot3}(a) also indicates that at this time CRs have created a huge low-density cavity at the center. 

The formation and final break-up of the cavity may be best seen in Figure \ref{plot4}. The top and middle panels show that the cavity is initially formed along the $z$ direction where CRs are injected, and then merges with the lower cavity ($z<0$; not shown in Figure \ref{plot4}) at the reflective boundary $z=0$. The cavity is formed as the CR pressure causes the expansion of the surrounding gas, driving a large gas outflow, as  indicated in Figure \ref{plot5}, which shows the time evolution of the gas mass $M_{\rm gas}(<r)$ enclosed within various radii $r=10,20,50,100$ kpc (see \citealt{guo10} for more discussions and implications). The outflow begins at small radii, and gradually affects gas at larger radii. The size (radius) of the cavity is very large, extending out to $80$-$100$ kpc at $t=100$ Myr. At later times, the cavity breaks up and thermal gas flows back to the cluster center as seen in the bottom panels of Figure \ref{plot4}. Figure \ref{plot5} suggests that the gas inflow happens during $t\sim 100$-$200$ Myr. The middle panels of Figure \ref{plot4} show a feature of gas inflow near the image centers, indicating that the gas indeed starts to flow inward as early as $t\sim100$ Myr, and the gas inflow is not spherically symmetric. The inflow of thermal gas also produces shocks, which propagate within the cavity and heat some low-density gas to very high entropy (a few hundred keV cm$^{2}$), as shown by the short-dashed line in Figure \ref{plot3}(c). We note that the amount of shock-heated gas within the cavity is very small and negligible compared to the inflowing thermal gas.

\begin{figure*}
\plotone{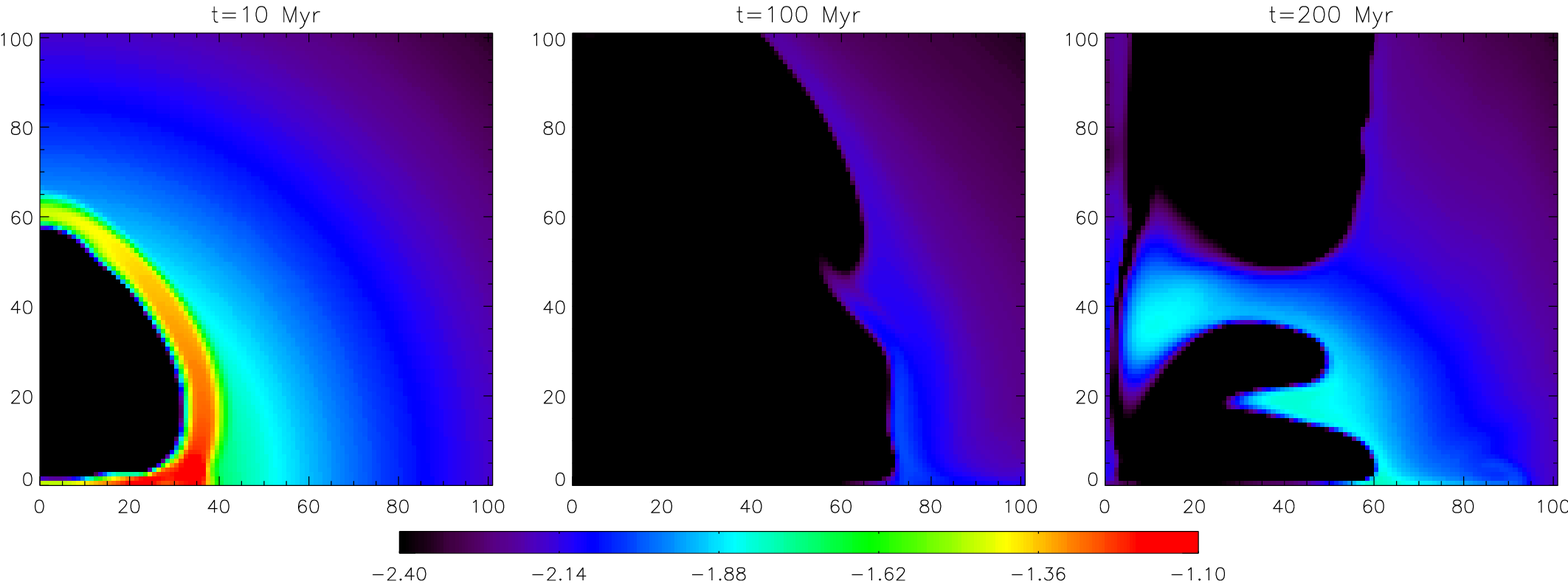}
\caption{Central slices ($100\times100$ kpc) of  log $(n_{\rm e}/{\rm cm}^{-3})$ at three time epochs t=$10$, $100$, $200$ Myr in run D1-A. The horizontal and vertical axes stand for $r$ and $z$ in kpc respectively. The images are saturated inside the cavities to show the variations of gas density in the ICM. The right image shows a ``Swiss-cheese-like" topology, which has been detected in the cluster Hydra A \citep{wise07}.}
 \label{plot7}
 \end{figure*}

The bottom panels of Figure \ref{plot4} also suggest that, as thermal gas flows back to the cluster center, the huge X-ray cavity is disrupted into two pairs of cavities in the orthogonal direction. This morphology is quite different from that of regular AGN feedback events, which usually produce one pair of X-ray cavities in opposite directions and are essential to maintain many clusters in the CC state. However in this paper, we investigate much more powerful AGN outbursts, which are much rarer and may transform CC clusters to the NCC state. Depending on where the CRs are deposited, these powerful AGN outbursts may produce X-ray cavities with different morphologies. In run D1-A, where the CRs are injected into the ICM continuously by a source moving out from the central AGN, the resulting central cavity is elongated in the jet direction and finally breaks up into three pairs of cavities in the same direction (see Figure \ref{plot7}). Interestingly, the morphology seen in this run may have already been detected in cluster observations (e.g., in Hydra A; see Figure \ref{plot7} and the corresponding discussion). 

As high-entropy thermal gas flows to the center, the cluster relaxes to the NCC state after $t\sim 0.2-0.3$ Gyr, as indicated by the long-dashed and dot-short dashed lines in Figure \ref{plot3}. During the NCC state, the cluster has a relatively flat temperature profile and a high entropy ($\gtrsim 80$ keV cm$^{2}$) core. The central cooling time in the NCC state is quite long ($\sim 3$ - $4$ Gyr) and the cluster profiles do not evolve much from $t=0.3$ to $0.5$ Gyr (dot-long dashed line). 

In our calculations, we ignore thermal conduction, which may remove irregularities in the temperature distribution and may even provide a strong heating source for the ICM during the CC to NCC transformation \citep{guo09,ruszkowski10,parrish10}. In cluster cool cores where the gas temperature decreases in the direction of gravity, conduction may be strongly suppressed by the heat flux driven buoyancy instability (hereafter HBI; see \citealt{quataert08}), which re-orients the magnetic field to be largely transverse to the radius. However, it is possible that AGN outbursts  counteract the HBI by disturbing the azimuthal nature of magnetic fields, thus enabling conduction and lowering the required AGN energy to accomplish the CC to NCC transformation. A detailed study of this process is clearly beyond the scope of this paper (but see \citealt{guo09} for a simplified one-dimensional study of this effect on galaxy groups and clusters). 

\begin{figure}
\plotone{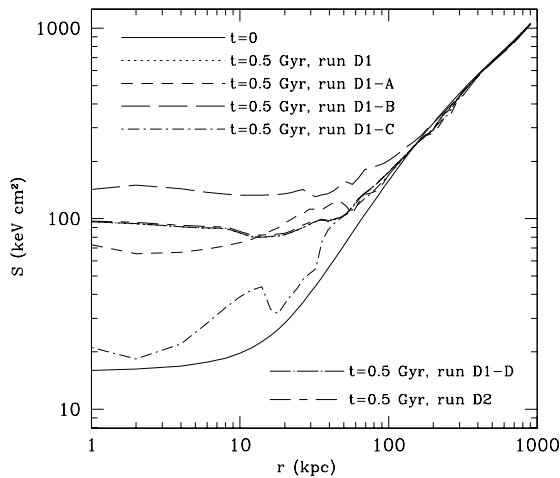}
\caption{Radial profiles of emission-weighted spherically averaged gas entropy at $t=0.5$ Gyr in various runs. The same initial profile for all our runs is also shown as the solid line for comparison.}
 \label{plot6}
 \end{figure}

We conducted a parameter study of our model by performing a series of simulations with varied parameters as listed in Table 1. The resulting radial profiles of emission-weighted spherically averaged gas entropy at $t=0.5$ Gyr in these runs are shown in Figure \ref{plot6}. In run D1-B, the injected AGN energy is doubled ($E_{\rm agn}=6.3\times 10^{61}$ erg), resulting in much higher core entropies at $t=0.5$ Gyr. Varying the initial abundance profile has a negligible effect on cluster evolution, as shown in runs D1, D1-D, and D2, all of which show a very similar entropy profile at $t=0.5$ Gyr. Obviously metallicity only affects the radiative cooling rate, which is not significant during the first one half Gyr in our simulations (shock heating increases the cooling time of thermal gas as well). 

However, the evolution of core entropy does depend on where the CRs are injected. When the CRs are injected at a larger radii $z_{\rm cav}=30$ kpc in run D1-C, the cluster can not be heated to the NCC state, as clearly shown by the dot-short dashed line in Figure \ref{plot6}. In this case, a pair of X-ray cavities are produced in opposite directions and do not merge at the cluster center as in run D1 (see \citealt{guo10} for more details of such a simulation in the cluster MS 0735.6+7421). Even when the injected CR energy is doubled, the gas within the central $\sim 10$-$20$ kpc still can not be heated to high entropies as in run D1. Since this very central region in CC clusters is usually quite dense, it is very likely that AGNs often deposit quite a large fraction of energy directly into this region. If so, powerful AGN outbursts can remove CCs as in run D1.
In run D1-A, we assume that the CRs are injected into the ICM continuously by a (jet) source moving out from the central AGN (more specifically, $z_{\rm cav}$ increases from $10$ to $50$ kpc at a constant speed within $t_{\rm agn}$). The resulting entropy profile at $t=0.5$ Gyr (the short-dashed line) clearly indicates that the cluster is heated to the NCC state, and, as expected, has higher entropies in the region $\sim10$-$50$ kpc than in run D1. It is of great interest to investigate what determines the position where AGN energy is deposited and forms X-ray cavities. Incorporating the relevant physics, this may be studied by high-resoluton jet simulations, which may also reveal some important processes that may affect the cavity evolution and that are not included in our current simulations (e.g., \citealt{sternberg07}; \citealt{sternberg08}).

Figure \ref{plot7} shows central slices of  log $(n_{\rm e}/{\rm cm}^{-3})$ at three epochs $t=10$, $100$, $200$ Myr in run D1-A. The injected CRs produce a low-density cavity, which is much more elongated in the $z$ direction than that in run D1. As the cavity breaks up, its morphology resembles three pairs of X-ray cavities, as seen in the right panel ($t=200$ Myr). Such multiple pairs of AGN bubbles have been detected in the cluster Hydra A (denoted as the ``Swiss-cheese-like" topology by \citealt{wise07}), which is currently hosting a very powerful $\sim 10^{61}$ erg AGN outburst, probably transforming that cluster to the NCC state. This also suggests that the multiple pairs of AGN bubbles in Hydra A are probably not an expression of the AGN duty cycle.

\subsection{Are Central Abundance Peaks also Removed?}
\label{section:rcap}

 \begin{figure}
\plotone{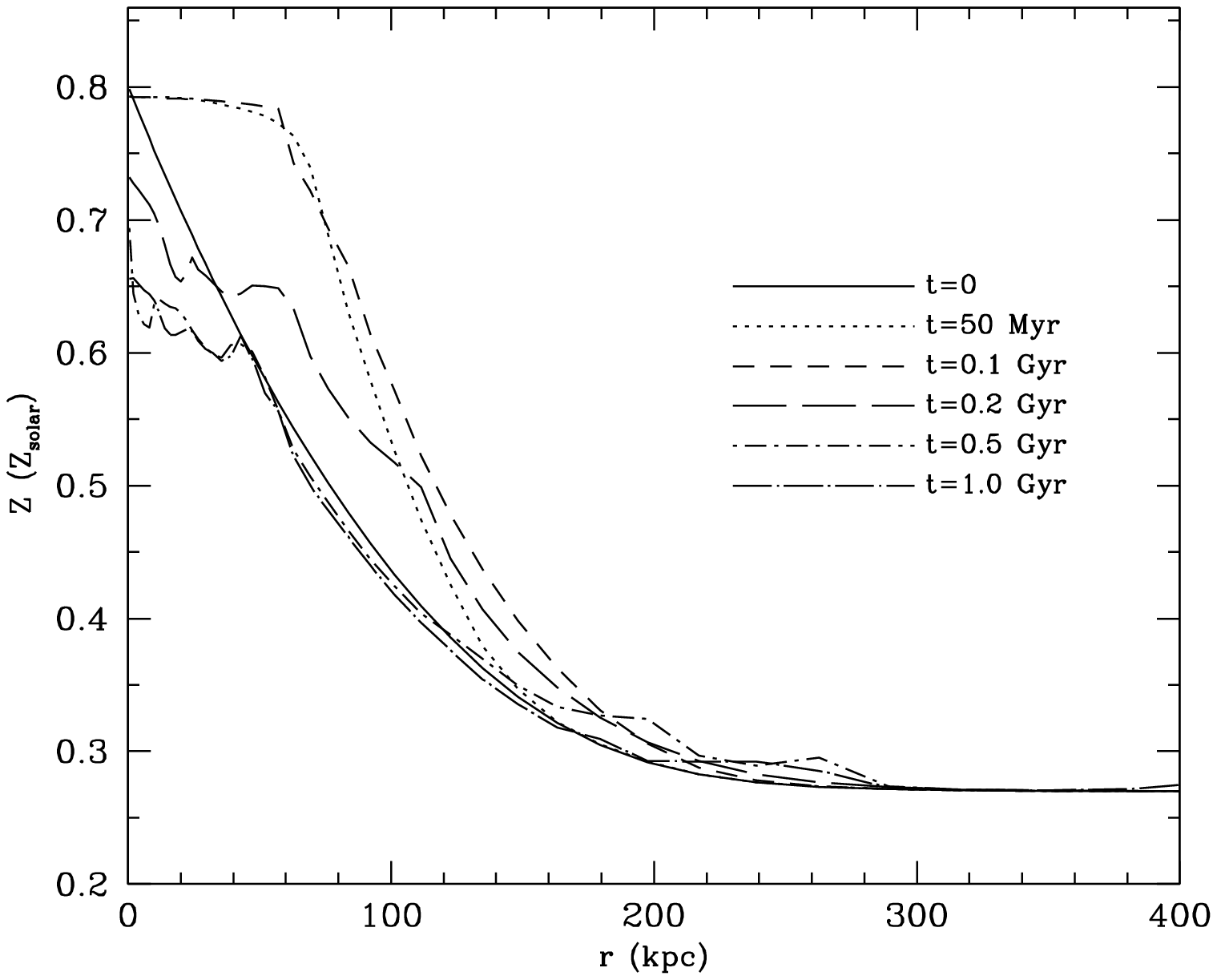}
\caption{Radial profiles of emission-weighted spherically averaged metallicity $Z$ at various times in our main run D1. In the final NCC state, the central abundance peak is not removed, though the peak abundance does decrease.}
 \label{plot8}
 \end{figure} 

 \begin{figure}
\plotone{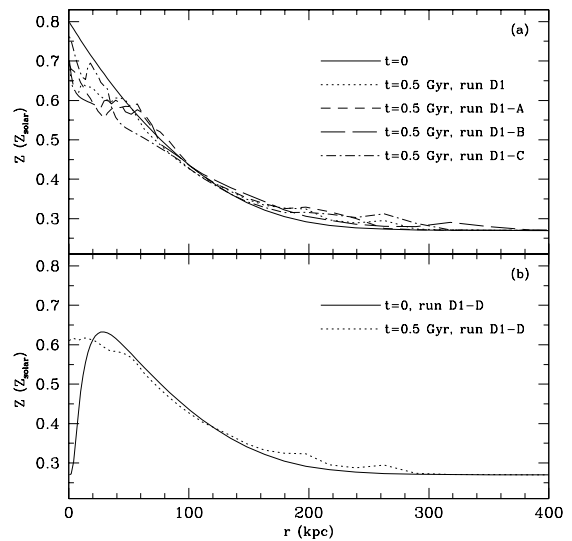}
\caption{Radial profiles of emission-weighted spherically averaged metallicity at $t=0.5$ Gyr in our runs (a) D1, D1-A, D1-B, D1-C, and (b) D1-D. The solid line in the top panel represents the same initial profile for runs D1, D1-A, D1-B, and D1-C, while that in the bottom panel shows the initial abundance profile in run D1-D. The central abundance peak is not removed in any of  these runs.}
 \label{plot9}
 \end{figure} 

 \begin{figure}
\plotone{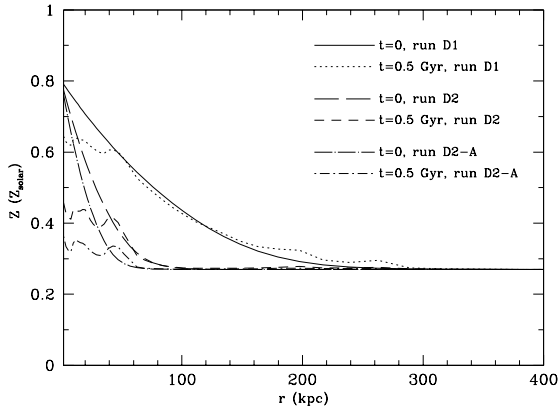}
\caption{Radial profiles of emission-weighted spherically averaged metallicity at $t=0$ and $t=0.5$ Gyr in our runs D1, D2, D2-A. The final NCC abundance profile depends sensitively on the peak size $r_{\rm metal}$ of the initial abundance profile. In runs D2 and D2-A, $r_{\rm metal}$ is comparable to or smaller than the radius ($\sim 50$-$60$ kpc for A1795) within which AGN outbursts can efficiently mix metals. For these runs (D2 and D2-A), the central abundance peaks are effectively removed, resulting in flat NCC abundance profiles which are often observed (e.g. \citealt{degrandi01}; \citealt{leccardi10}).}
 \label{plot10}
 \end{figure} 

Heavy metals observed in the hot ICM using X-ray spectroscopy indicate that in CC clusters the iron abundance profiles have central peaks and decline radially outward \citep{degrandi01, degrandi04, baldi07,leccardi08}. Interestingly, observations reach contradictory results on the abundance profiles of NCC clusters. It has been claimed by \citet{degrandi01} and \citet{degrandi04} that NCC clusters have a nearly uniform spatial distribution of metals. In contrast, observations by \citet{leccardi08} and \citet{sanderson09} indicate that the abundance profiles of NCC clusters are very similar to those of CC clusters, both showing central peaks. More recently, \citet{leccardi10} found both types of NCC clusters in their cluster sample. In this subsection, we investigate how the centrally-peaked abundance profile evolve when the CC cluster is transformed to the NCC state by AGN outbursts. We construct a physical scenario that can  explain both observationally contradictory results on the NCC abundance profiles.
 
We first investigate the temporal evolution of the abundance profile in run D1. Figure \ref{plot8} shows radial profiles of emission-weighted spherically averaged metallicity $Z$ at various time epochs in this simulation. Clearly during the early times (the dotted and short-dashed lines) as the low-density cavity is created at the cluster center, the abundance profile is shifted toward larger radii, while the metallicity within the cavity is nearly constant. As the cavity breaks up and the cluster relaxes to the NCC state, high-metallicity gas moves back to the cluster core.  As clearly seen in Figure \ref{plot8} (the dot-short dashed and dot-long dashed lines), the gas abundance within the central $\sim 50$ kpc in the NCC state is efficiently diluted by mixing, while the abundance profile at large radii remains similar to that in the original CC state. However, since the size of abundance peak in the original CC abundance profile is very broad ($r_{\rm metal}=160$ kpc), the abundance profile in the final NCC state still has a central peak, with a maximum abundance value of around $0.6-0.65$, which is roughly the abundance value of the initial CC profile at a cluster-centric radius of $50$ kpc. Thus we find that in this run, though the abundance peak value does decrease, the central abundance peak is not removed during the CC to NCC transformation, which is consistent with recent observations by \citet{leccardi08} and \citet{sanderson09}, while not explaining NCC clusters with relatively flat abundance profiles observed by \citet{degrandi01} and \citet{degrandi04}.
   
Figure \ref{plot9}(a) shows radial profiles of emission-weighted spherically averaged metallicity at $t=0.5$ Gyr in our runs D1, D1-A, D1-B, and D1-C. Clearly the central abundance peak is not removed in any of these runs, suggesting that it is very difficult to remove it by varying the energy or location of the CR injection. X-ray observations indicate that some clusters and elliptical galaxies exhibit a dip in abundance in their very centers, e.g., Abell 2199 (see the long-dashed line in Figure \ref{plot2}). We performed another run D1-D to study if such a CC abundance profile may result in a final NCC abundance profile without the central abundance peak. To implement a dip in abundance profile at the very center, we introduce an inner cutoff term to $Z_{\rm r}$ in Equation (\ref{metalr}):
\begin{eqnarray}
Z_{\rm r}=0.8e^{-r/r_{\rm metal}}(1-e^{-r/r_{0}})
\text{,}  \label{metalr2} 
\end{eqnarray} 
where $r_{0}=10$ kpc is a characteristic inner cutoff radius. The initial abundance profile in run D1-D is then calculated using Equation (\ref{metaleq}), and is shown as the solid line in Figure \ref{plot9}(b). The profile has a peak abundance of $\sim 0.6$-$0.65$ at a cluster-centric radius of $\sim 20$-$30$ kpc, which is roughly the same radius where the peak in the abundance profile of Abell 2199 is located. The dotted line in Figure \ref{plot9}(b) shows the spherically averaged radial abundance profile at $t=0.5$ Gyr when the cluster is in the NCC state, clearly indicating that the abundance peak is not removed, though the metals within $\sim 50$ kpc are efficiently mixed. This result is reasonable, since the total amount of gas within the central abundance dip is relatively small (the gas mass within the central $20$ kpc is roughly one tenth of that within the central $50$ kpc).

We performed runs D2 and D2-A to explore the dependence of the final NCC abundance profile on the peak size $r_{\rm metal}$ of the initial CC abundance profile. This is motivated by the fact that in some clusters $r_{\rm metal}$ is much smaller (e.g., $r_{\rm metal}\sim 40$-$60$ kpc in A2199, as seen in Figure \ref{plot2}) than we assumed in our previous runs. The results are shown in Figure \ref{plot10}. For run D2 with $r_{\rm metal}=60$ kpc, the final NCC abundance profile is relatively flat and has a central value of around $0.4$ in the cluster central regions, which is consistent with observations by \citet{degrandi01}, \citet{degrandi04}, and \citet{baldi07}. In run D2-A with a smaller peak size $r_{\rm metal}=40$ kpc, the resulting NCC abundance profile is even flatter with a smaller central value of around $0.3$. 

From Figure \ref{plot10}, we conclude that the final NCC abundance profile depends sensitively on the peak size $r_{\rm metal}$ of the initial CC abundance profile. It can be effectively removed when $r_{\rm metal}$ is comparable to or smaller than a characteristic radius ($\sim 50$-$60$ kpc $\sim 0.03 r_{200}$ for A1795) within which powerful AGN outbursts can efficiently mix metals. Interestingly, X-ray observations find two contradictory (possibly bimodal) types of NCC clusters according to their central abundance profiles: NCC clusters with \citep{leccardi08, sanderson09} and without central abundance peaks \citep{degrandi01, degrandi04}. Our analysis can thus naturally explain both types of NCC clusters, which may be formed by powerful AGN outbursts from CC clusters with central abundance profiles having different spatial sizes.
 
\subsection{Implications on the Evolution History of Galaxy Clusters}
\label{section:implication}
 
X-ray observations suggest that the iron excess associated with central abundance peaks is mainly produced by the brightest cluster galaxies (BCGs; \citealt{degrandi04}; \citealt{bohringer04}) and the enrichment times are usually quite long ($\gtrsim 5$ Gyr; \citealt{bohringer04}). In the absence of mixing, the metallicity profiles should follow the optical light profiles of the BCGs and have narrow central abundance peaks, easily destroyed when strong AGN outbursts transform CC clusters to the NCC state. However, recent observational and theoretical studies suggest that CC clusters are maintained in the CC state by numerous AGN feedback events, which also gradually broaden central abundance peaks by generating turbulent mixing \citep{rebusco05,rebusco06} or gas outflows \citep{guo10}. If a cluster stays in the CC state for a very long duration, it would build up a broad central abundance peak, which is very robust to the CC to NCC transformation. If our explanation for the origin of NCC clusters is correct, it implies that the two types of NCC clusters have different evolution histories. NCC clusters without central abundance peaks were never in a CC state for a very long duration, while those with central abundance peaks could have experienced a long CC state.

\begin{figure}
\plotone{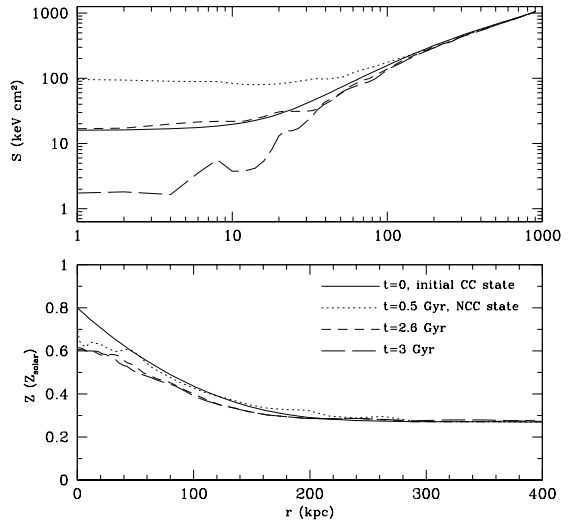}
\caption{Time evolution of radial profiles of emission-weighted spherically averaged gas entropy (top panel) and metallicity (bottom panel) in run D1. Due to the powerful AGN outburst, the cluster reaches the NCC state at $t\sim 0.2-0.3$ Gyr. As the ICM cools at later times, it settles inward gradually. However, the resulting decrease in the size of the central abundance peak is not large ($r_{\rm metal}\sim 130$ kpc at $t\sim 3$ Gyr).}
 \label{plot11}
 \end{figure}

NCC clusters may cool gradually and reach the CC state. During this transformation, the cluster gas slightly settles inward, which may affect metallicity profiles. In particular, for NCC clusters with central abundance peaks, the sizes of abundance peaks may decrease. We study this effect by following run D1 for much longer times. Figure \ref{plot11} shows time evolution of radial profiles of emission-weighted spherically averaged gas entropy and metallicity in this run. At $t\sim 2.6$ Gyr, the cluster reaches a state very close to the initial CC state, and the abundance profile does slightly shift toward the cluster center. However, the decrease in the spatial size of the abundance peak is small, and $r_{\rm metal}$ ($\sim 130$ kpc) at $t\sim 2.6$ Gyr is much larger than the characteristic radius ($\sim 50$-$60$ kpc) within which powerful AGN outbursts can efficiently remove the metallicity peak. At $t\sim 3$ Gyr, the cluster core cools to much lower entropies, but the abundance profile does not change appreciably compared to that at $t\sim 2.6$ Gyr, retaining a broad central abundance peak. This is reasonable, since during the NCC to CC transformation, the increase of gas mass within $\sim 60$ kpc is less than one tenth of the gas mass within the original NCC metallicity peak. Consequently another powerful AGN outburst later on could not remove this broad abundance peak. Furthermore, simulations by \citet{poole08} indicate that broad central abundance peaks can not be easily destroyed by mergers. Therefore, broad central abundance peaks, once formed, may persist for the remaining life of clusters. 

Some low-redshift CC clusters have narrow central abundance peaks (e.g., A2199). These systems may have stayed in the CC state for a relatively short duration, otherwise a broad central abundance peak would have been established by AGN feedback. Interestingly, current X-ray observations seem to indicate that most low-redshift ($z \lesssim 0.1$; \citealt{degrandi04}) and intermediate-redshift ($0.1\lesssim z \lesssim 0.3$; \citealt{baldi07}; \citealt{leccardi08}) CC clusters have quite broad central abundance peaks ($\sim 0.1$-$0.2 r_{\rm 200}$). This may suggest that most low-redshift NCC clusters without central abundance peaks were formed at higher redshift, when more CC clusters had smaller $r_{\rm metal}$. Observers often emphasize mean metallicity profiles averaged over many clusters (e.g., \citealt{degrandi04}; \citealt{leccardi08}), but our analysis suggests that it may be informative to statistically study the distribution of $r_{\rm metal}$ in a large sample of CC clusters (and NCC clusters as well). An analysis of temporal evolution of the averaged $r_{\rm metal}$ over cosmic time is also of great interest. The relative fraction of both types of NCC clusters and its redshift evolution may reveal the evolution history of the ICM thermodynamics and AGN activity. 

Alternatively, recent cosmological simulations by \citet{burns08} and \citet{henning09} suggest that the origins of CC and NCC clusters depend on their different merger histories: CC clusters accreted mass slowly over time and grew enhanced CCs via hierarchical mergers, while NCC clusters experienced major mergers early in their evolution that destroyed embryonic CCs and produced conditions that prevent CC reformation. However, as we discussed in Section 1, current cosmological simulations do not have enough resolution to accurately model core structure and evolution. Furthermore, many NCC clusters have a short cooling time ($\sim 1$ Gyr; \citealt{sanderson06}), and it is unclear why radiative cooling should not change their thermal state if they are formed early in their evolution history. Similarly, this scenario could not explain the origin of NCC clusters with flat metallicity profiles; central abundance peaks would be established by long-time metal enrichment from the BCGs if powerful AGN outbursts are not invoked to destroy them.

\section{Discussion and Summary}
\label{section:conclusion}

X-ray observations of galaxy clusters indicate a striking bimodality in the properties of cluster cores, which separate clusters into two distinct classes: CC and NCC clusters. The origin of this bimodality remains unclear. At the same time, a handful of extremely powerful, core-changing AGN outbursts have recently been detected in clusters with a total energy $\sim 10^{61}-10^{62}$ erg \citep{nulsen05a, nulsen05, mcnamara05}. By conducting a suite of two-dimensional axisymmetric hydrodynamical simulations, we inquire if such strong AGN outbursts can transform a CC cluster to the NCC state, and if so, how this transformation happens. Cluster abundance profiles have been well observed by X-ray observations and contain important information on physical processes in clusters. Consequently, we also follow the evolution of iron abundance profiles and study how they are affected by AGN outbursts during the CC to NCC transformation.

We assume that AGN energy is injected into the ICM mainly in the form of CRs, and follow the co-evolution of these CRs  with the cluster gas in our simulations. We find that when a large fraction of this energy is injected near the cluster center, strong AGN outbursts with energy $\sim 10^{61}-10^{62}$ erg (see Table 1) can completely remove cool cores, transforming the CC cluster (A1795 as our fiducial cluster) to the NCC state. In view of the high central gas density in CC clusters, deposition of AGN energy within the cluster core may be common. The deposition of CRs produces shocks in the ICM, which propagate outward and heat the core gas to high entropies ($\sim 100$ kev cm$^{2}$) very quickly (within $\sim 10$-$50$ Myr in our main run D1). The CRs also drive a strong gas outflow, producing a large, low-density cavity near the cluster center (with a radius of $\sim 80-100$ kpc), which breaks up at time $t\sim 200$ Myr. During the break-up of the cavity, high-entropy thermal gas flows back to the cluster center and is efficiently mixed. The cluster relaxes to the NCC state with a central entropy $\sim 100$ kev cm$^{2}$ by time $t\sim 200$-$300$ Myr.

Recent X-ray observations reveal several extremely powerful AGN outbursts with energy $\sim 10^{61}-10^{62}$ erg, including Hydra A \citep{nulsen05a}, Hercules A \citep{nulsen05} and MS0735.6+7421 (\citealt{mcnamara05}; \citealt{mcnamara09}). MS0735.6+7421 hosts the most powerful AGN outburst currently known with an estimated AGN energy $\sim 10^{62}$ erg, but most of the energy seems to be deposited beyond $40$ kpc (see \citealt{guo10} for a detailed modeling of this outburst), and the CC in this cluster, though significantly heated, is not completely removed. In contrary, the temperature profile in Hydra A is quite flat, suggesting that the CC in this cluster may have been removed by the observed powerful AGN outburst. Furthermore, the observed ``Swiss-cheese-like" topology associated with three pairs of X-ray cavities in Hydra A \citep{wise07} is also well reproduced in our run D1-A, where CRs are injected into the ICM continuously by a (jet) source moving out from the central AGN. Hercules A hosts a young ($t\sim 59$ Myr) AGN outburst ($\sim 3 \times 10^{61}$ erg), which has  significantly disturbed the cluster core \citep{nulsen05}. While Hercules A is not fully relaxed, there is already some evidence suggesting that it may be a NCC cluster \citep{white97}. Deep X-ray observations are needed to reveal its detailed thermal structure.

During the transformation of CC to NCC systems, AGN outbursts efficiently mix metals in cluster central regions (within $\sim 50-60$ kpc $\sim 0.03 r_{200}$ for A1795), where abundance peaks are usually seen in CC systems. The maximum abundance in the central peak thus decreases, but the peak is not removed if its spatial size ($r_{\rm metal}$) is large. However, when $r_{\rm metal}$ is comparable to or smaller than the characteristic radius within which AGN outbursts can efficiently mix metals (as in the cluster A2199), the central peak is effectively removed and the resulting NCC abundance profile is relatively flat. Our model naturally explains these two distinct types of NCC clusters seen in observations -- those 
 with and without central abundance peaks -- both of which can develop following powerful AGN outbursts in CC clusters with different initial spatial abundance peak sizes. 
 
In contrast, the merger scenario for creating NCC clusters may not explain those NCC systems without strong central abundance peaks. Simulations by \citet{poole08} indicate that merging clusters which initially host central abundance peaks do not yield merger remnants with flat metallicity profiles (but also see \citealt{vazza10}), though the dependence of their results on the size of the initial abundance peak remains to be explored. A statistical analysis of the peak sizes and their redshift evolution combined with a study of the possibly redshift-dependent relative fraction of both types of NCC clusters may shed interesting insights on the evolution history of the ICM thermal state and AGN activity in galaxy clusters.

\acknowledgements

We thank the anonymous referee for a detailed and helpful report, which improved the presentation of the paper. Studies of the evolution of hot cluster gas at UC Santa Cruz are supported by NSF and NASA grants for which we are very grateful.

\bibliography{ms} 
    
\label{lastpage}

\end{document}